\documentstyle[twoside,fleqn,espcrc2,psfig]{article}
\newcommand{\be}{\begin{equation}}
\newcommand{\ee}{\end{equation}}
\newcommand{\bea}{\begin{eqnarray}}
\newcommand{\eea}{\end{eqnarray}}
\newcommand{\no}{\noindent}
\newcommand{\nn}{\nonumber}

\newcommand{\Tr}{{\rm Tr\,}}
\newcommand{\Det}{{\rm Det\,}}
\newcommand{\e}{{\rm e\,}}

\title{$1/M$ correction to quenched QCD with non-zero baryon density}
\author{
Gert Aarts$^{1}$,
Olaf Kaczmarek$^{2}$,
Frithjof Karsch$^{2}$
and Ion-Olimpiu Stamatescu$^{1,3}$\\
[5mm]
{\small $^1$Institut f\"ur Theoretische Physik, Univ. Heidelberg, Heidelberg, Germany}\\
{\small $^2$Fakult\"at f\"ur Physik, Univ. Bielefeld,  Bielefeld, Germany\ \ \ \ }
{\small $^3$FESt,  Heidelberg, Germany}
}
\begin{document}
\begin{abstract}We study the $\kappa^2$ corrections to the 
quenched limit of $\mu>0$ QCD. We use an improved reweighting procedure.
\end{abstract}
\maketitle
\begin{figure}[htb]
\vspace{4cm} 
\center{
\includegraphics{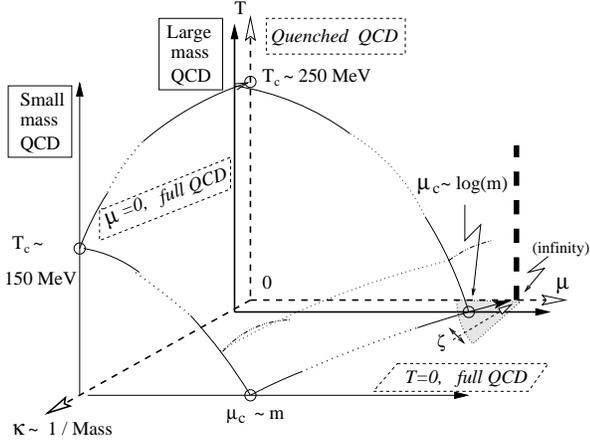}}
\caption{
Tentative phase diagram.}
\vspace{-0.6cm} 
\label{f.diag}
\end{figure}

\vspace{-1cm}
\no {\bf Introduction}\ \ 
In Fig.\ \ref{f.diag} we sketch tentative features of the QCD phase
diagram in the $\kappa,\, \mu,\, T$ space. Since a general description is
missing, one may try to concentrate on special situations, such as the
``quenched" limit \cite{fktre} (see below), represented by the thick
vertical dashed line at infinite $\mu$ in the $\kappa=0$ plane. For a more
physical situation we consider a vertical plane at small but non-zero
$\kappa$, then, e.g., a possible transition would appear at
$\mu_c(\kappa,T) < \infty$. Here we present a first attempt to describe
phenomena away from the quenched limit.

\no {\bf Hopping parameter expansion}\ \
The QCD grand canonical partition function
is (we specify for one flavour of Wilson quarks
and neglect constant factors; $U$ are  links, $T$ lattice translations):
\bea
{\cal Z}(\beta,\kappa,\mu) = \int[DU]\, 
\e^{-S_G(\beta,\{U\})}{\cal Z}_F({ {\kappa}}, \mu, \{U\}), \nn 
\eea 
\vspace*{-0.6cm}%
\bea
{\cal Z}_F({ {\kappa}}, \mu, \{U\}) =  
\Det W ({ {\kappa}}, \mu, \{U\}),
\eea
\vspace*{-0.8cm}%
\bea 
W =\!\!\!\!\!\! && \!\!\!\! 1 -  \kappa\, \sum_{i=1}^3 \left( 
\Gamma_{+i}\,U_i\,T_i +
\Gamma_{-i}\,T^*_i\,U^*_i\right) \nn \\
\!\!\!\!\!\! && \!\!\!\!
-\kappa\,  \left( \e^{\mu}\,\Gamma_{+4}\,U_4\,T_4 +
\e^{-\mu}\,\Gamma_{-4}\,T^*_4\,U^*_4 \right), 
\label{e.act}
\eea
with $\Gamma_{\pm \mu} = 1 \pm \gamma_{\mu}$, $2\kappa = 
\frac{1}{{\tilde m}+3+\cosh \mu} = \frac{1}{m+4}$.
Here ${\tilde m}$ is the naive
 continuum limit ``bare mass", $m$ the bare mass at $\mu=0$. The analytic expansion in
$\kappa$ leads to an expansion in closed loops, the links contributing 
factors $\kappa$ or $\kappa \e^{\pm \mu}$, see (\ref{e.act}):
\bea
&& \!\!\!\! {\cal Z}_F({ {\kappa}}, \mu, \{U\}) = \Det W = {\rm exp} ( \Tr 
\ln W ) \nn \\
&& \!\!\!\!  =  {\rm exp} \left[-\sum_{l=1}^\infty  \sum_{\left\{{\cal
C}_l\right\}} \sum_{s=1}^\infty ~{{{ (\kappa^l f_{{\cal C}_l})}^s}\over
s}\,\Tr_{\rm D,C}     {\cal L}_{{\cal C}_l}^s \right] \nn \\
&& \!\!\!\! = 
\prod_{l=1}^{\infty} ~\prod_{\left\{{\cal C}_l\right\}}~ 
  \Det_{\rm D,C} \left(1~-~{ {\kappa}} ^l f_{{\cal C}_l}
{\cal L}_{{\cal C}_l}\right)   \label{e.hopg}
\eea
\no (D, C $=$ Dirac, colour). ${\cal C}_l$ are distinguishable, 
non-exactly-self-repeating 
closed paths of length $l$, $s$ is the number of times
a loop ${\cal L}_{{\cal C}_l}$ covers  ${\cal C}_l$,
\bea
f_{{\cal C}_l} = \left(\epsilon \, \e^{\pm N_{\tau}\mu}\right)^r\ {\rm 
if}\ 
{\cal C}_l = ``Polyakov\,\, r\!\! - \!\! path" \!,\! \label{e.fact}
\eea
\no 1 otherwise. A $``Polyakov\ r\!\! - \!\! path"$  
closes over the lattice 
in the $\pm 4$  direction with winding number $r$ and
periodic(antiperiodic) b.c.\ [$\epsilon = +1(-1)$]. 

\no {\bf Quenched limit at $\mu>0$}\ \ In the double limit \cite{bend}
$\kappa \rightarrow 0,\, \mu \rightarrow \infty$ with 
$\kappa\, \e^{\mu} \equiv \zeta$ fixed,
only straight, positive Polyakov loops ${\cal P}_{\vec x}$ are retained:
\bea
{\cal Z}_F^{[0]}(C, 
\left\{U\right\}) = \exp \left[- 
  \sum_{\left\{{\vec x}\right\}}
\sum_{s=1}^\infty\!\! ~{{{ (\epsilon C)}^s}\over s}~\Tr_{\rm C} 
  ({\cal P}_{\vec x})^s \right] \nn
\eea
\vspace{-0.5cm}
\bea
=\, \prod_{\left\{{\vec x}\right\}}~ 
  \Det_{\rm C} \left(1~-~\epsilon\,C {\cal P}_{\vec x}\right)^2,
\,\,\,\,\,\, C = (2\, \zeta)^{N_\tau} \label{e.hdl} 
\eea
\no (see \cite{bend}, and \cite{fktre,bhtky} for further studies).
 $C$ (or $\zeta$) fixes the direction toward the limit - see Fig.\ 1.

\no {\bf Next order corrections}\ \ 
To disentangle the effects of 
$\kappa$ and $\mu$ we go to next order in $\kappa$:
\bea
{\cal Z}_F^{[2]}({ {\kappa}}, \mu, \left\{U\right\}) 
 =   {\rm exp}\left\{-2\,  \sum_{\left\{{\vec x}\right\}}\,
\sum_{s=1}^\infty \,{{{ (\epsilon\, C
)}^s}\over s} \right. {\times} \nn 
\eea
\vspace{-0.4cm}
\bea
\left. \Tr_{\rm C} 
  \left[({\cal P}_{\vec x})^s  + \kappa^2\sum_{r,q,i,t,t'}
(\epsilon\, C)^{s(r-1)}({\cal P}_{{\vec x},i,t,t'}^{r,q})^s \right]\right\}
\nn
\eea
\vspace{-0.4cm}
\bea
=&& \!\!\!\!\!\!\!\!   {\cal Z}_F^{[0]}( C,  \left\{U\right\}) \times \nn 
\\
&& \!\!\!\!\!\!  \prod_{{\vec x}, r,q,i,t,t'} 
  \Det_{\rm C} \left(1-(\epsilon\,C)^{r}\,\kappa^2\,
  {\cal P}_{{\vec x},{i},t,t'}^{r,q}\right)^2 . 
\label{e.corr2}
\eea
For easy bookkeeping we use the temporal gauge
$U_{n,4}=1$, except for $U_{({\vec x}, n_4=N_\tau),4} \equiv V_{\vec x}$:
free, and
\be
{\cal P}_{{\vec x},i,t,t'}^{r,q} = (V_{\vec x})^{r-q} 
U_{({\vec x},t),i} (V_{{\vec x}+{\hat {\i}}})^q 
U_{({\vec x},t'),i}^*
\ee
\no with
$r>q\ge 0$, $i =\pm 1, \pm 2, \pm 3$, 
$1\le t \le t' \le N_{\tau}$ ($t < t'$ for $q=0$).
As an approximation to the full theory one may expect (\ref{e.corr2}) to
be good in some small region around the vertical limit line and by varying
$\mu$ or $T$ we may hope to get information, say, about the phases around
$\mu_c$ at large mass. Alternatively we may propose (\ref{e.corr2}) as a
model by itself. At large $\mu$ and mass it would be an approximation in
the above sense at all $T$. At large $T$ it would also be an approximation
for any $\mu$, as we recover the ``Polyakov loop model" there \cite{fkpol}. We
may thus hope that our model retains some features of  QCD with full
$\kappa$, $\mu$ dependence.  

\begin{figure}[htb]
\vspace{5.2cm} 
\center{
\includegraphics{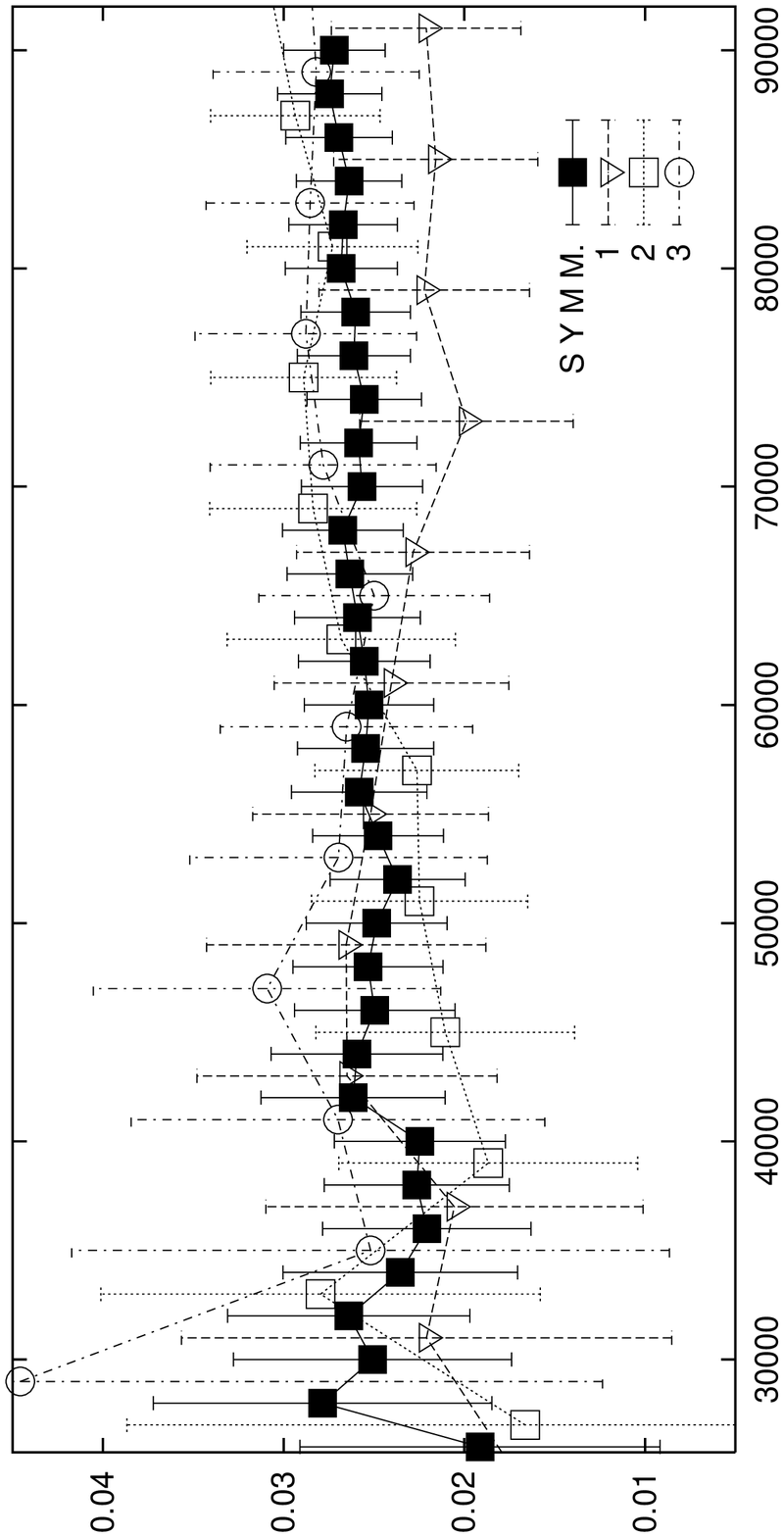}
\includegraphics{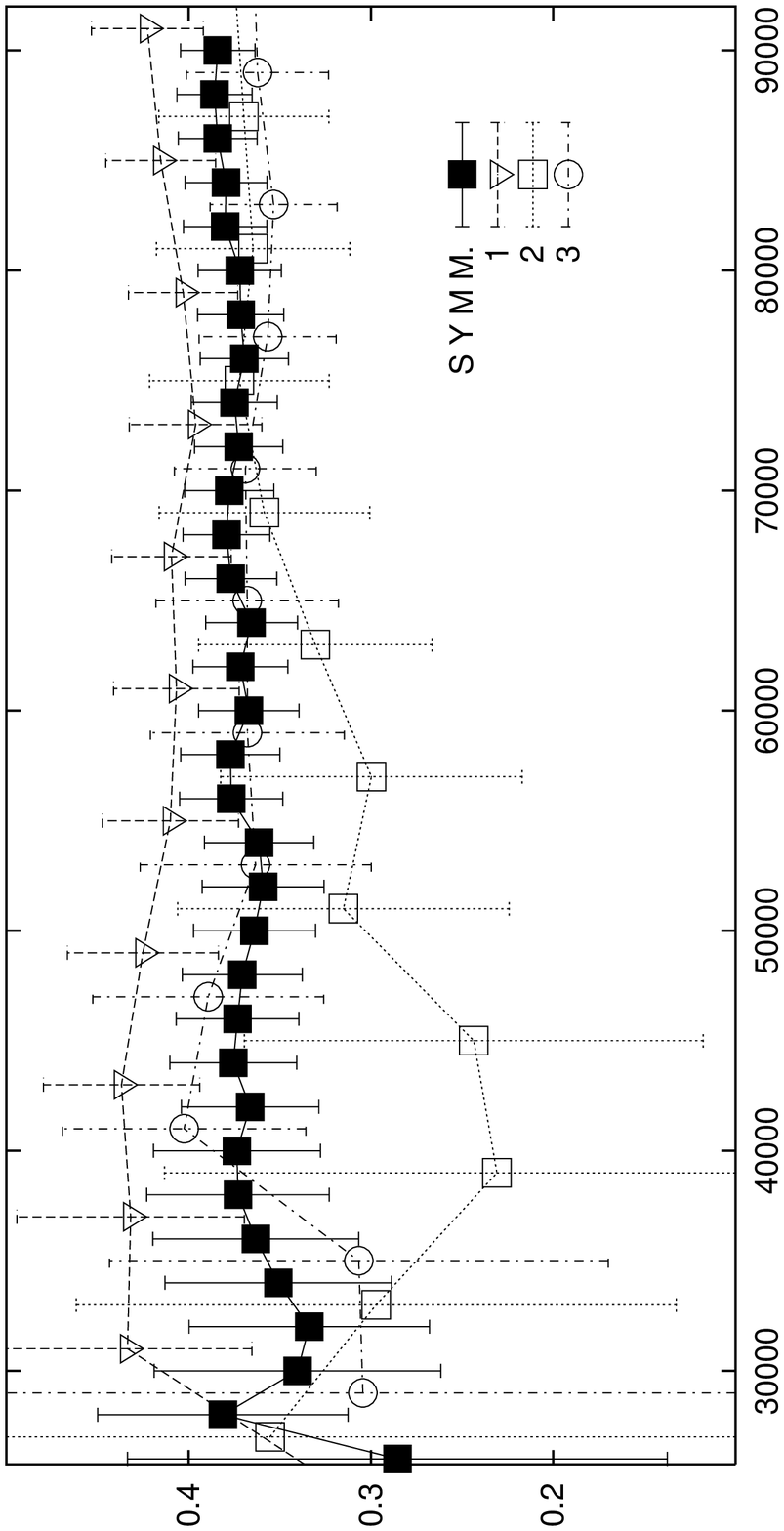}} 
\caption{History of weight (above) and density (below),
$N_{\tau}=4,\,\beta=5.5,\,\kappa=0.12,\,\mu=0.6267$.} 
\vspace{-0.5cm} 
\label{f.conv} 
\end{figure}

\no {\bf Improving  convergence of $\mu>0$ simulations}\ \
Since at $\mu>0$ ${\cal Z}_F$ is complex, simulations using reweighting
methods suffer from a ``sign problem".  In this situation one can try to
realize the cancellations between contributions in a correlated way.  
This can be done by identifying appropriate transformations of the
configurations and correspondingly separate the measure into symmetric and
antisymmetric factors, to be used in the MC generation and reweighting,
respectively. In special cases this method can be developed to a solution
of the sign problem \cite{alf}, in our case we may hope to achieve some
improvement - see also \cite{phl}. The sensitive dependence of ${\cal
Z}_F$ on the temporal Polyakov loops suggests to use center
rotations. This is especially suggestive for our model
(\ref{e.hdl}-\ref{e.corr2}), but the method can be applied also to the
full problem. We write
\bea
{\cal Z} \!\!\! &=& \!\!\!\int[DU\,DV]
\, \e^{-S_G(\{U,V\})}{\cal Z}_F( \{U,V\}) 
\nn \\
\!\!\!&=& \!\!\! \int[DU\,DV]
\, {\cal S}(\{U,V\}){\cal T}(\{U,V\}) \label{e.sym}
\eea
\no with center-symmetric ${\cal S}(\{U,V\}) = {\cal S}(\{U,ZV\})$,
$Z = \left\{1, \e^{2\pi i /3}, \e^{4\pi i /3}\right\}$, and calculate 
\bea
\langle O \rangle = 
\frac{\langle\sum_{Z} O^Z{\cal T}^Z\rangle_{\cal S}}
{\langle\sum_{Z}{\cal T}^Z\rangle_{\cal S}},\ \ 
X^Z = X(\{U,ZV\}). \label{e.ave}
\eea
\no 
Here $\langle\cdot \rangle_{\cal S}$ are averages taken with the 
Boltzmann  factor ${\cal S}$. $U$ denote spatial links. We choose
\bea
{\cal S} = \e^{-S_G}\,\left|\sum_{Z} {\cal Z}_F( \{U,ZV\})\right|. 
\label{e.sb1}
\eea
Alternatively, in the approximation of retaining only the 
$s=r=1$ terms in (\ref{e.corr2}) we can take for ${\cal S}$
\bea
\e^{-S_G - 2\epsilon C\left|
\, \sum_{\left\{{\vec x}\right\}}
\Tr_{\rm C} 
  \left[{\cal P}_{\vec x} \, +\, \kappa^2\,\sum_{\pm {i},t,t'}
{\cal P}_{{\vec x},{i},t,t'}^{1,0} \right]\right|}. \label{e.sb2}
\eea
Note that ${\cal S}$ already includes $\kappa$ and $\mu$. 
In Fig.\ \ref{f.conv} we show the convergence
of the weight ${\cal T}$ and  charge density averages (\ref{e.dens}) for 
each $Z$ in (\ref{e.ave}) and after symmetrization. Note the
good convergence in a regime of large cancellations ($C=0.0407$; the 
weight is only $\sim 3\%$). The symmetrization improves the statistics by  
a factor 3-10 in the confining phase. In the deconfining phase, whichever
sector is chosen by ${\cal S}$ the method automatically captures the right
contribution. 
For the analysis we use $8^34$ and $8^36$ lattices with $\epsilon = 
-1$, performing $\sim 10^5$ heat-bath/over-relaxation/metropolis sweeps 
with (\ref{e.sb1}) or (\ref{e.sb2}) and $r=1$ (first 20000 sweeps for 
thermalization). Errors are estimated with a jackknife analysis. 

\begin{figure}[ht]
\vspace{6cm} 
\center{
\includegraphics{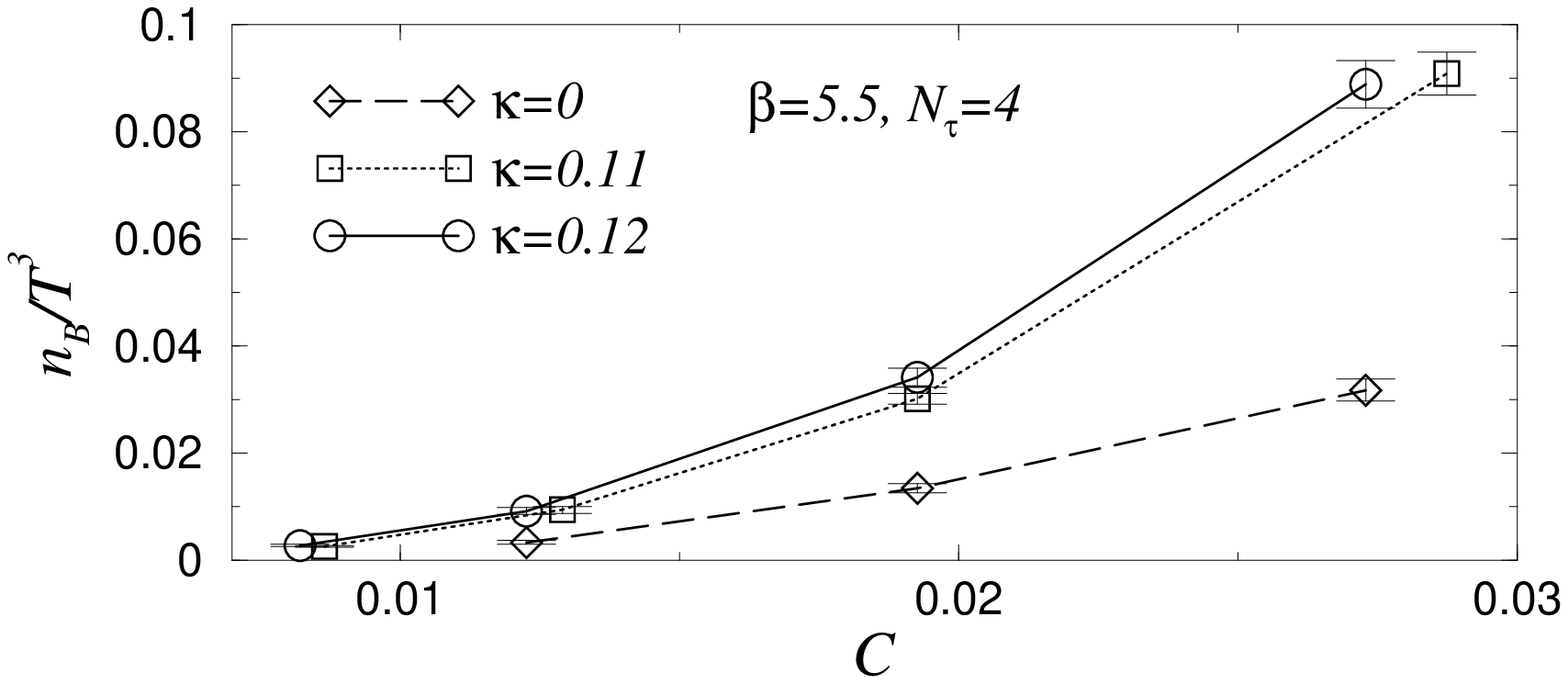}
\includegraphics{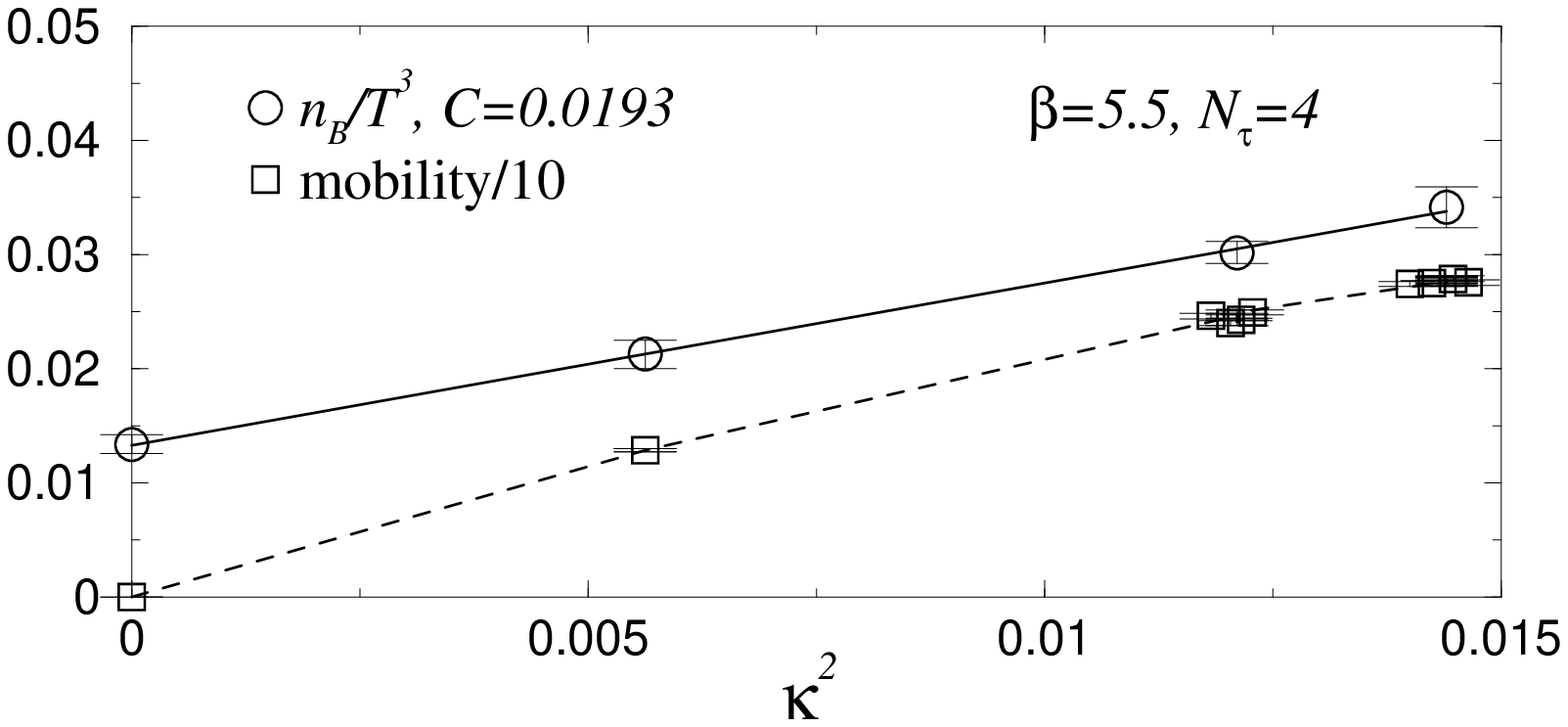}} 
\caption{Above: $n_B /T^3$ vs $C$, various $\kappa$. Below: $n_B 
/T^3$ at fixed $C$ and $m_B/10$ vs $\kappa^2$.} 
\label{f.kappa}
\vspace{-0.5cm} 
\end{figure}
 
\begin{figure}[\protect h]
\vspace{6cm} 
\center{
\includegraphics{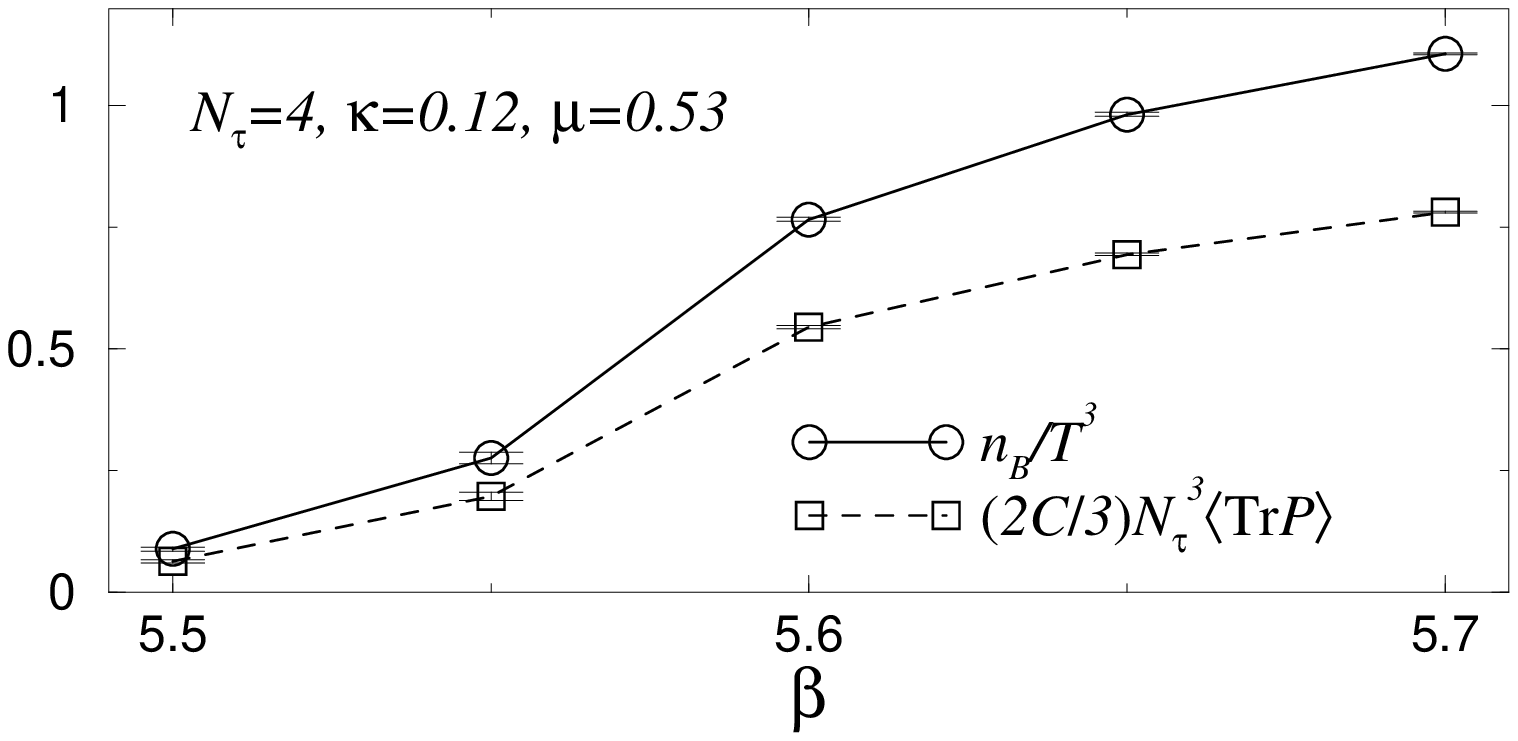} 
\includegraphics{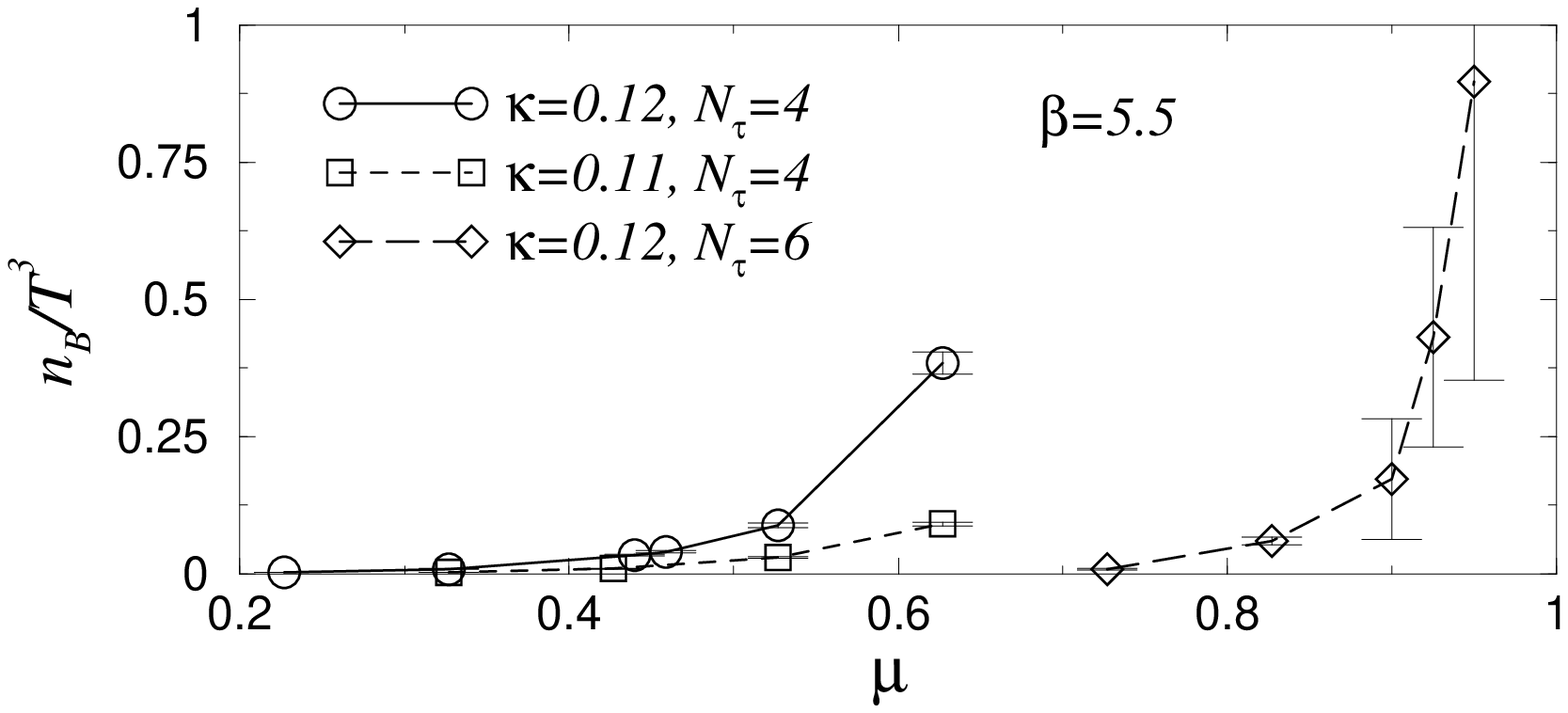}} 
\caption{ Above: $n_B/T^3$ and Polyakov loop
vs $\beta$. Below: $n_B/T^3$ vs $\mu$.}
\vspace{-0.5cm} 
\label{f.mu}
\end{figure}

\no {\bf Results}\ \ 
We calculate the baryon number density $n_B$
and the ``mobility" $m_B$ (we introduce this as a normalized measure for 
the effect of the $\kappa^2$ term):
\vspace{-0.4cm}
\bea
\label{e.dens}
\frac{n_B}{T^3} \!\!\! &=& \!\!\! \frac{N_{\tau}^3}{3 N_{\sigma}^3}\, ({ 
\hat n_0 +\hat n_1}), \,\,\,\,\,\,
m_B = \frac{\hat n_1}{\hat n_0 +\hat n_1}, \\
\hat n_0 \!\!\!&=&\!\!\!  2 C \langle \sum_{\vec x} \Tr P_{\vec x} 
\rangle,\\
\hat n_1 \!\!\!&=&\!\!\!  2C {\kappa}^2 \langle
\sum_{\vec x, \pm {i},t,t'} \Tr P_{{\vec x}, {i},t,t'} \rangle,
\eea
as well as $\langle \left(\sum_{\vec x} \Tr P_{\vec x}\right)^3 \rangle$, 
the
spatial and temporal plaquettes and the topological susceptibility.
The effect of the $\kappa^2$ correction can be
observed in Fig.\ \ref{f.kappa}. As expected,  the mobility   
depends primarily on $\kappa$, not on $\mu$, and
the correction to the density depends linearly on $\kappa^2$ 
and it can be 
significant. The most important point, however, is that we can now 
probe the $\kappa$ and $\mu$  directions independently. 
The question of phase transitions is investigated in Fig.\ \ref{f.mu}, 
both 
along the $T$ and along the $\mu$ direction. We 
observe  the installation of the deconfining regime with increasing $T$ 
(the temporal Polyakov loop is non-zero in the confining phase due to the 
fermions). Also,
a signal for a transition in $\mu$ is seen, especially at small temperature.
We cannot say, however, from these preliminary results whether these are
genuine phase transitions or mere crossovers. The further development 
 implies a more detailed analysis.

\no {\bf Acknowledgments:} We are indebted to M. Alford, J. Berges and C.
Wetterich for discussions.

\end{document}